\documentclass[usenatbib]{mn2e}

\usepackage{graphicx}

\begin{document}

\title[Effect of a M.V.F on a M.F.D]{Analysis of the Effect of a Mean
Velocity Field on Mean Field Dynamo}
\author[A. Kandus]{Alejandra Kandus \thanks{%
email: kandus@uesc.br} \\
Laborat\'{o}rio de Astrof\'{\i}sica Te\'{o}rica e Observacional,
Departamento de Ci\^{e}ncias Exatas e Tecnol\'{o}gicas, \\
UniversidadeEstadual de Santa Cruz, Rodov\'{\i}a Ilh\'{e}us-Itabuna km 15
s/n, CEP: 45662-000, Salobrinho, Ilh\'{e}us, BA, Brazil}
\maketitle

\begin{abstract}
We study semi-analytically and in a consistent manner, the generation of a
mean velocity field $\overline{\mathbf{U}}$ by helical MHD turbulence, and the
effect that this field can have on a Mean Field Dynamo. Assuming a
prescribed, maximally helical small scale velocity field, we show that large
scale flows can be generated in MHD turbulent flows, via small scale Lorentz
force. These flows back-react on the mean electromotive force of a Mean Field
Dynamo through new terms, leaving the original $\alpha $ and $\beta $ terms
explicitly unmodified. Cross-helicity plays the key role in interconnecting
all the effects. In the minimal $\tau$ closure that we chose to work with,
the effects are stronger for large relaxation times.
\end{abstract}


\section{Introduction}

Magnetohydrodynamical (MHD) turbulence seems to be a major physical process
to generate and maintain the magnetic fields observed in most of the
structures of the Universe \citep{bran-sub-rpt,zeldovich83}. When addressing
the problem of the generation of large scale magnetic fields by small scale
turbulent flows, a model known as Mean Field Dynamo (MFD) is usually
considered \citep{moffat}. Despite its simplicity and lack of broad
applicability, it proved to be a very useful tool in studying qualitatively
conceptual issues of large scale magnetic field generation. The mechanism is
based on decomposing the fields into large scale, or mean fields, 
$\overline{\mathbf{U}}$, $\overline{\mathbf{B}}$, $\overline{\mathbf{A}}$ and
small scale, turbulent ones $\mathbf{u}$, $\mathbf{b}$, $\mathbf{a}$. These
small scale fields have very small coherence length, but their intensities
can be higher than the one of the mean fields. In this mechanism the
evolution equation for $\overline{\mathbf{B}}$ is written $\partial 
\overline{\mathbf{B}}/\partial t=\nabla \times \left( \overline{\mathbf{U}}%
\times \overline{\mathbf{B}}+\overline{\mathbf{\mathcal{E}}}-\eta \overline{%
\mathbf{J}}\right) $, where $\overline{\mathbf{J}}=\nabla \times \overline{%
\mathbf{B}}$, $\eta $ is the Ohmic resistivity and $\overline{\mathbf{\mathcal{E%
}}}=\overline{\mathbf{u}\times \mathbf{\ b}}$ the turbulent electromotive force
(t.e.m.f.) \footnote{Overlines denote local spatial averages: they represent 
vector quantities
whose intensities may vary in space, but whose direction and sense are
uniform or vary smoothly. $\left\langle {}\right\rangle _{vol}$ denote
volume averages, i.e., quantities that can depend only on time.}. The term $%
\overline{\mathbf{U}}\times \overline{\mathbf{B}}$ is usually disregarded in
the studies of MFD as the focus of most of them is to understand the
generation of large scale quantities due to small scale effects. If
homogeneous and isotropic turbulence is considered, the t.e.m.f can be
written as $\overline{\mathbf{\mathcal{E}}}=\alpha \overline{\mathbf{B}}%
-\beta \overline{\mathbf{J}}$, with $\alpha \simeq -\left( 1/3\right) \tau
_{corr}\left[ \overline{\mathbf{u}\cdot \left( \nabla \times \mathbf{u}%
\right) }-\overline{\mathbf{b}\cdot \left( \nabla \times \mathbf{b}\right) }%
\right] $ and $\beta \simeq \left( 1/3\right) \tau _{corr}\overline{u^{2}}$, 
$\tau _{corr}$ being a correlation time 
\citep{moffat72,rudiger74,pouquet76,zeldovich83}. The dependencies on $%
\overline{\mathbf{B}}$ and $\mathbf{b}$ are due to the back-reaction of
those induced fields on the dynamo \citep{moffat,bran-sub-rpt}. In the
kinematically driven dynamo considered here, the features
of the generated fields crucially depend on the helicity of the flows:
helical flows are at the base of the mechanisms to generate large scale
fields, while non-helical flows would only produce small scale fields. This
separation, however, is somewhat artificial, as small scale fields are also
produced by helical turbulence \citep{bran-sub-rpt}.

In this paper we want to address an issue not (or very seldom) considered in
the literature up to now, namely, the induction of large scale flows 
$\overline{\mathbf{U}}$, also named shear flows, by the small scale turbulent
fields, and how these induced flows back-react on the turbulent
electromotive force $\overline{ \mathbf{\mathcal{E}}}$ of a MFD. On one side, 
we mean that the expression 
$\partial \overline{\mathbf{B}}/\partial t=\nabla \times \left( \overline{%
\mathbf{\mathcal{E}}}-\eta \overline{\mathbf{J}}\right) $ would be valid
only during the time interval in which $\overline{\mathbf{U}}\times 
\overline{\mathbf{B}}\ll \overline{\mathbf{\mathcal{E}}}$; and on the other,
even if this conditions is satisfied, $\overline{\mathbf{\mathcal{E}}}$
could be affected by the generation of $\overline{\mathbf{U}}$ and
consequently its functional form should be modified to incorporate this
effect. The generation of magnetic fields due to the action of these large
scale velocity flows instead of by $\overline{\mathbf{\mathcal{E}}}$ was
recently addressed analytically by several authors %
\citep{roga-kleo-03,roga-kleo-04,radler-stepa-05}, and was also studied
numerically by \citet{brandenburg01} and semi-analytically by 
\citet{bb-2002}. However, none of those works addressed specifically the 
issue we want to analyze here.

We work in the framework of the two scale approximation, that consists in
assuming that mean fields peak at a scale $k_{L}^{-1}$ while turbulent ones
do so at $k_{S}^{-1}\ll k_{L}^{-1}$, and also consider homogenous and
isotropic turbulence. Although this kind of turbulence is of dubious
validity when dealing with large scale fields, it serves well for initial,
qualitative studies of the sought effects. Another assumption we shall make
is that $\overline{\mathbf{B}}$ is force-free, i.e., of maximal current
helicity. Although fields with this feature can be observed in certain
astrophysical environments, they are not a generality, and also they are not
seen in some numerical simulations. The main reason to use them here is to
simplify the (heavy) mathematics, while maintaining a physically meaning
scenario.

In order to find $\overline{\mathbf{\mathcal{E}}}$ when Lorentz force acts
on the plasma, we must solve a differential equation that contains terms
with one point triple correlations, i.e., averages of products of three
stochastic fields evaluated at the same point. This means that instead of
dealing with only one equation to solve for $\overline{\mathbf{\mathcal{E}}}$%
, we have to solve a hierarchy of them. In order to break this hierarchy and
thus simplify the mathematical treatment of the problem we must choose a 
\emph{closure prescription}, which consists in writing the high order 
correlations as functions of the lower order ones, but maintaining the physical 
features of the problem under study. In MHD the intensity of the non-linearities 
is measured by the magnetic Reynolds number, which is defined from the
induction equation for the magnetic field, $\partial \mathbf{B}/\partial t=%
\mathbf{\nabla }\left( \mathbf{U}\times \mathbf{B}\right) +\eta \nabla ^{2}%
\mathbf{B}$, as $R_{m}\sim \left( UB/l\right) /\left( \eta B/l^{2}\right)
=Ul/\eta $, where now $U$ is a characteristic velocity of the plasma, $l$ is
a characteristic length and $\eta $ is the ohmic resistivity. 
Thus for $R_{m}\ll 1$ the non-linear terms can be neglected in front of the
resistive ones, and therefore the equations become linear, while for $%
R_{m}\gg 1$ the resistive terms should be dropped off in front of the
non-linear ones. The intermediate regime is more difficult to analyze. In
this paper we shall consider $R_{m}\gg 1$, i.e., the non-linearities must be
maintained, and consequently a closure scheme must be selected to deal with
them. We choose to work with the so called minimal $\tau $ approximation 
\citep{bf2002,bran-sub-rpt}, whereby the triple moments in the equation for 
$\overline{\mathbf{\mathcal{E}}}$ will be considered as proportional to 
quadratic moments, and written in
the form $\zeta \overline{\mathbf{\mathcal{E}}}$, with the proportionality
factor $\zeta \sim \tau _{rel}^{-1}$, where $\tau _{rel}$ is a relaxation
time that can in principle be scale, and/or $R_m$, dependent. The validity 
of this closure was checked numerically \citep{bran-sub-aa} for low Reynolds 
numbers, and was also verified for the case of passive
scalar diffusion. We shall assume that it is also valid for all the triple
correlations that may appear throughout this work. We consider boundary
conditions such that all total divergencies vanish. These conditions may be
a bit unrealistic for astrophysical systems, but they have two advantages:
magnetic helicity becomes a gauge-invariant quantity \citep{berger84}, and
the obtained results can be compared with numerical simulation, as to
perform them it is customary to use those conditions. We consider fully
helical, prescribed, $\mathbf{u}$ fields.

Our starting points are the evolution equations for $\overline{\mathbf{%
\mathcal{E}}}$ and for the magnetic helicities $H_{L,S}^{M}$, as the
evolution of these quantities is tightly interlinked %
\citep{bf2002,bran-sub-rpt} in the absence of shear flows. When including $%
\overline{\mathbf{U}}$ new terms appear in the equation for $\overline{%
\mathbf{\mathcal{E}}}$, but the ones that drive the evolution of $\overline{%
\mathbf{\mathcal{E}}}$ in absence of $\overline{\mathbf{U}}$, i.e., the $%
\alpha $ and $\beta $ terms \citep{moffat72,rudiger74,pouquet76,bf2002}, are
not explicitly modified. For those new terms, further equations must be
derived, that in turn show the subtleties of the interplay among $\overline{%
\mathbf{U}}$, $\mathbf{u}$, $\mathbf{b}$ and $\overline{\mathbf{B}}$. Due to
the chosen boundary conditions, the evolution equations for $H_{L}^{M}$ and $%
H_{s}^{M}$ do not explicitly depend on $\overline{\mathbf{U}}$, they will do
so implicitly through $\overline{\mathbf{\mathcal{E}}}$. We consider fully
helical $\overline{\mathbf{U}}$ fields, so we study their growth through its
associated kinetic helicity, $H^{\overline{U}}\equiv \left\langle \left( 
\mathbf{\nabla }\times \overline{\mathbf{U}}\right) \cdot \overline{\mathbf{U%
}}\right\rangle _{vol}$ and show that, for fully helical, prescribed $%
\mathbf{u}$, large scale flows will be always generated, as long as small
scale Lorentz force is not null, i.e., if $\overline{\left( \mathbf{\nabla }%
\times \mathbf{b}\right) \times \mathbf{b}}\not=0$. We shall consider two
values for the magnetic Reynolds number, $R_{m}=200$ and $2000$, and for
each case analyze the effect of short and large $\tau _{rel}$. In general we
find that for short $\tau _{rel}$ (large $\zeta $), i.e., strong
non-linearities,  the effect of large scale flow is negligible, thus
producing results that practically do not differ from the ones in the
absence of large scale flows. For large $\tau _{rel}$ (small $\zeta$) 
the general effect is an enhancement of the electromotive force and the 
inverse cascade of magnetic helicity, this enhancement being stronger for 
$R_{m}=2000$ than for $R_{m}=200$.

\section{Main Equations}

Ohm's law for an electrically conducting fluid reads $\mathbf{E}=-\mathbf{U}
\times \mathbf{B}+\eta \mathbf{J}$, with $\eta $ the electric resistivity
and $\mathbf{J}$ the electric current. The equation for $\mathbf{B}$ is the
induction equation:

\begin{equation}
\frac{\partial \mathbf{B}}{\partial t}=-\mathbf{\nabla }\times \mathbf{E,}
\label{c}
\end{equation}
and from $\mathbf{B}=\mathbf{\nabla }\times \mathbf{A}$ we have

\begin{equation}
\mathbf{E}=-\mathbf{\nabla }\Phi -\frac{\partial \mathbf{A}}{\partial t},
\label{b}
\end{equation}
which is an evolution equation for $\mathbf{A}$. The equation for the
velocity field $\mathbf{U}$ is the Navier-Stokes equation that, when
considering only Lorentz force, reads

\begin{equation}
\frac{\partial \mathbf{U}}{\partial t}=-\left( \mathbf{U}\cdot \mathbf{\
\nabla }\right) \mathbf{U}-\frac{\nabla p}{\rho }+\left( \mathbf{\nabla }
\times \mathbf{B}\right) \times \mathbf{B}
-\nu \mathbf{\nabla }\times \left( 
\mathbf{\nabla }\times \mathbf{U}\right) .  \label{d}
\end{equation}
with $\nu $ the kinetic viscosity. To work within mean field theory 
\citep{moffat} we decompose the different fields as $\mathbf{B}=\overline{
\mathbf{B}}+ \mathbf{b}$, $\mathbf{A}=\overline{\mathbf{A}}+\mathbf{a}$, 
$\mathbf{U}=\overline{\mathbf{U}} +\mathbf{u}$, $\mathbf{E}=\overline{\mathbf{
E}}+\mathbf{e}$ and $\mathbf{\Phi }= \overline{\mathbf{\Phi }}+\mathbf{\phi }
$, where any mean value of stochastic quantities vanishes. The derivation of
the evolution equations for the mean and stochastic fields is a standard
procedure, already described in the literature \citep{zeldovich83,bf2002}.
Consequently we only write here the results. Assuming incompressibility of
the large and small scale flows, considering that $\overline{\mathbf{B}}$ is
force free and working with Coulomb gauge for the vector potential, i.e., 
$\mathbf{\nabla }\cdot \overline{\mathbf{A}} =0=\mathbf{\nabla }\cdot \mathbf{
a}$, we obtain the following equations for the mean fields
\begin{equation}
\frac{\partial \overline{\mathbf{B}}}{\partial t}=\mathbf{\nabla }\times %
\left[ \overline{\mathbf{U}}\times \overline{\mathbf{B}}+\overline{\mathbf{%
\mathcal{E}}}-\eta \mathbf{\ \nabla }\times \overline{\mathbf{B}}\right] ,
\label{g}
\end{equation}
\begin{equation}
\frac{\partial \overline{\mathbf{A}}}{\partial t}=\bar{\mathbf{P}}\left[ 
\overline{\mathbf{U}}\times \overline{\mathbf{B}}+\overline{\mathbf{\mathcal{
E}}}\right] -\eta \mathbf{\nabla }\times \overline{\mathbf{B}},  \label{h3}
\end{equation}
where $\overline{\mathbf{\mathcal{E}}}= \overline{\mathbf{u}\times \mathbf{b}}$ 
is the t.e.m.f., and 
\begin{eqnarray}
\frac{\partial \overline{\mathbf{U}}}{\partial t}&=&\bar{\mathbf{P}}\left[
-\left( \overline{\mathbf{U}}\cdot \nabla \right) \overline{\mathbf{U}}-
\overline{\left( \mathbf{u}\cdot \nabla \right) \mathbf{u}}+\overline{\left( 
\mathbf{b}\cdot \nabla \right) \mathbf{b}}\right] \nonumber\\
&+&\nu \nabla ^{2}\overline{\mathbf{U}}.  \label{i2}
\end{eqnarray}
$\left( \bar{\mathbf{P}}\right) _{ij}=\delta _{ij}\partial ^{2}-\partial
_{i}\partial _{j}$ is the projector that selects the subspace of solutions
of eq. (\ref{b}) that satisfy the Coulomb gauge condition and the subspace
of solutions of (\ref{d} ) that satisfy the incompressibility condition.
Observe that eq. (\ref{i2}) shows that a large scale velocity field can be
induced from an initially zero value, as long as $-\overline{\left( \mathbf{u%
}\cdot \nabla \right) \mathbf{u}}+\overline{\left( \mathbf{b}\cdot \nabla
\right) \mathbf{\ b}}\neq 0$. The equations for the small scale fields read

\begin{equation}
\frac{\partial \mathbf{b}}{\partial t}=\nabla \times \left[ \overline{%
\mathbf{U}}\times \mathbf{b}+\mathbf{u}\times \overline{\mathbf{B}}+\mathbf{u%
}\times \mathbf{b}-\overline{\mathbf{\mathcal{E}}}\right] +\eta \nabla ^{2}%
\mathbf{b},  \label{j}
\end{equation}

\begin{equation}
\frac{\partial \mathbf{a}}{\partial t}=\bar{\mathbf{P}}\left[ \overline{%
\mathbf{U}}\times \mathbf{b}+\mathbf{u}\times \overline{\mathbf{B}}+\mathbf{u%
}\times \mathbf{b}-\overline{\mathbf{\mathcal{E}}}\right] +\eta \nabla ^{2}%
\mathbf{a},  \label{k2}
\end{equation}%
and 
\begin{eqnarray}
\frac{\partial \mathbf{u}}{\partial t} &=&\bar{\mathbf{P}}\left[ -\left( 
\overline{\mathbf{U}}\cdot \nabla \right) \mathbf{u}-\left( \mathbf{u}\cdot
\nabla \right) \overline{\mathbf{U}}-\left( \mathbf{u}\cdot \nabla \right) 
\mathbf{u}+\overline{\left( \mathbf{u}\cdot \nabla \right) \mathbf{u}}\right.
\nonumber \\
&+&\left. \left( \overline{\mathbf{B}}\cdot \nabla \right) \mathbf{b}+\left( 
\mathbf{b}\cdot \nabla \right) \overline{\mathbf{B}}+\left( \mathbf{b}\cdot
\nabla \right) \mathbf{b}-\overline{\left( \mathbf{b}\cdot \nabla \right) 
\mathbf{b}}\right] \nonumber\\
&+&\nu \nabla ^{2}\mathbf{u}.  \label{l2}
\end{eqnarray}

\subsection{Evolution Equation for Derived Quantities: Magnetic Helicity,
Large Scale Kinetic Helicites, and the Stochastic Electromotive Force}

As stated in the Introduction, we want to study if an initially zero, or
very weak $\overline{\mathbf{U}}$, can grow due the action of a MFD, and
back-react on it and on the magnetic helicity. This last quantity is defined
as the average over the entire volume of the dot product $\mathbf{A}\cdot 
\mathbf{B}$ \citep{biskamp}. In this way, we write the magnetic helicity
associated to the large and small scale fields respectively as $%
H_{L}^{M}\equiv \left\langle \overline{\mathbf{A}}\cdot \overline{\mathbf{B}}%
\right\rangle _{vol}$ and $H_{S}^{M}\equiv \left\langle \mathbf{a}\cdot 
\mathbf{b}\right\rangle _{vol}$, and by definition they can only depend on
time. The evolution equations for $H_{L}^{M}$ and $H_{s}^{M}$, for the
chosen boundary conditions, read \citep{bf2002} 
\begin{equation}
\frac{\partial H_{L}^{M}}{\partial t}=2\left\langle \overline{\mathbf{%
\mathcal{E}}}\cdot \overline{\mathbf{B}}\right\rangle _{vol}-2\eta
\left\langle \left( \nabla \times \overline{\mathbf{B}}\right) \cdot 
\overline{\mathbf{B}}\right\rangle _{vol},  \label{n}
\end{equation}%
and 
\begin{equation}
\frac{\partial H_{S}^{M}}{\partial t}=-2\left\langle \overline{\mathbf{%
\mathcal{E}}}\cdot \overline{\mathbf{B}}\right\rangle _{vol}-2\eta
\left\langle \left( \nabla \times \mathbf{b}\right) \cdot \mathbf{b}%
\right\rangle _{vol}.  \label{p}
\end{equation}%
Observe that these equations have the same form as the ones obtained in the
absence of large scale flows. This fact is due to the selected boundary
conditions: magnetic helicity can be injected into the system through the
boundaries by large scale flows. Thus, in the case under consideration here,
these flows cannot explicitly transport magnetic helicity between the
different scales, they will act implicitly through $\overline{\mathbf{%
\mathcal{E}}}$.

From the definition of $\overline{\mathbf{\mathcal{E}}}$ given above, the
evolution equation for the t.e.m.f. is $\partial \overline{\mathbf{\mathcal{E%
}}}/\partial t=\overline{\left( \partial \mathbf{u/}\partial t\right) \times 
\mathbf{b}}+\overline{\mathbf{u}\times \left( \partial \mathbf{b/}\partial
t\right) }$. Proceeding in a similar form as in Refs. \citep{bf2002,yo}, it
now reads 
\begin{eqnarray}
\frac{\partial \overline{\mathbf{\mathcal{E}}}}{\partial t} &=&\frac{1}{3}
\nabla \cdot \overline{\mathcal{E}}~ \overline{\mathbf{U}}+
\frac{2}{3}\overline{\mathbf{u}\cdot \mathbf{b}}\left( \nabla \times 
\overline{\mathbf{U}}\right)  \label{r} \\
&&+\frac{1}{3}\left[ \overline{\left( \mathbf{\nabla }\times \mathbf{b}
\right) \cdot \mathbf{b}}-\overline{\mathbf{u}\cdot \left( \mathbf{\nabla }
\times \mathbf{u}\right) }\right] \overline{\mathbf{B}}  \nonumber \\
&&-\frac{1}{3}\overline{u^{2}}\left( \mathbf{\nabla }\times \overline{
\mathbf{B}}\right) +\eta \overline{\mathbf{u}\times \nabla ^{2}\mathbf{b}}
+\nu \overline{\nabla ^{2}\mathbf{u}\times \mathbf{b}}+\overline{\mathbf{T}}.
\nonumber
\end{eqnarray}
where $\overline{\mathbf{T}}$ are the triple correlations for which a closure
must be applied.
Observe that the presence of $\overline{\mathbf{U}}$ adds two new terms to
the equation for $\overline{\mathbf{\mathcal{E}}}$ but does not explicitly
modify the ones found in the absence of those flows. The influence of $%
\overline{\mathbf{U}}$ on the terms proportional to $\overline{\mathbf{B}}$
will be through the dependence of those terms with the magnetic helicities
(cf. Ref. \citet{bf2002}). To gain conceptual clearness we shall make
further physical hypotheses on our systems, that will also help to simplify
the mathematics. One of them is to consider that large scale flows 
$\overline{\mathbf{U}}$ are fully helical. This is consistent with the
concept of mean field dynamo and with the chosen boundary conditions.
Therefore to track the evolution of the large scale velocity flow, we shall
study its associated kinetic helicity, defined as $H^{\overline{U}}
=\left\langle \overline{\mathbf{W}}\cdot \overline{\mathbf{U}}\right\rangle
_{vol}$, where $\overline{\mathbf{W}}=\nabla \times \overline{\mathbf{U}}$
is the vorticity. The derivation of the equation for $H^{\overline{U}}$ is
explained in the Appendix, and the result is 
\begin{equation}
\frac{\partial H^{\overline{U}}}{\partial t}\simeq 2\overline{\left( \mathbf{%
\nabla }\times \mathbf{b}\right) \times \mathbf{b}}\cdot \overline{\mathbf{W}%
}+2\nu \nabla ^{2}\overline{\mathbf{U}}\cdot \overline{\mathbf{W}}.
\label{t}
\end{equation}%
where the semi-equality is due to the fact that we are approximating volume
average by a local spatial average. It is well known that kinetic helicity
is not conserved in MHD \citep{biskamp} , so eq. (\ref{t}) is not an
essentially new result. However it serves to our purposes in showing that
large scale helical flows can be induced by turbulent $\mathbf{b}$-fields,
provided they are not force-free. At this point we make another supposition: 
we take $\mathbf{\nabla }\cdot \overline{\mathbf{\mathcal{E}}}=0$ which, 
besides being consistent with the chosen boundary conditions, means that 
the induction of large scale
magnetic fields maximal for $\overline{\mathbf{U}}=0$ (cf. eq. \ref{g} ).
Observe that by imposing Coulomb gauge on eq. (\ref{h3})\footnote[2]{
We cannot use equation (8) for $\mathbf{a}$ to calculate $\langle \mathbf{%
\nabla }\cdot \mathbf{\mathcal{E}}\rangle _{0}$ as it is trivially zero.} we
obtain a further constraint on the mean fields, namely $\mathbf{\nabla }%
\cdot \overline{\mathbf{\mathcal{E}}}=-\overline{\nabla \cdot \left( 
\overline{\mathbf{U}}\times \overline{\mathbf{B}}\right) }$, and the fact
that we consider it equal to zero allows us to replace $\overline{\mathbf{B}}%
\cdot \left( \mathbf{\nabla }\times \overline{\mathbf{U}}\right) =\overline{%
\mathbf{U}}\cdot \left( \mathbf{\nabla }\times \overline{\mathbf{B}}\right)$.

\section{Implementing the two scale approximation}

As was advanced in the Introduction, we shall work within the two scale
approximation, whereby mean fields are supposed to peak at a scale $%
k_{L}^{-1}$, and stochastic ones at $k_{S}^{-1}\ll k_{L}^{-1}$. We begin by
noting that eq. (\ref{r}) together with the definition of the t.e.m.f. and
eq. (\ref{t}), are very complicated, involving new functions of the mean and
stochastic fields for which further equations must be deduced. From the 
constraint on $\overline{\mathbf{U}}$ and $\overline{\mathbf{B}}$ derived
from $\nabla \cdot \overline{\mathbf{\mathcal E}}=0$, and the fact that we
are considering $\overline{\mathbf B}$ as force-free, we can write
$\overline{\mathbf U}\cdot \left(\nabla \times \overline{\mathbf{B}}\right)
\simeq k_{L}\overline{\mathbf{U}}\cdot \overline{\mathbf{B}}$, where the
last semi-equality stems from the fact that $\overline{\mathbf{B}}$ is
considered to be force-free. From eqs. (\ref{n}) and (\ref{p}), we see that
the dot product of $\overline{\mathbf{\mathcal{E}}}$ with $\overline{\mathbf{%
B}}$ is responsible for magnetic helicity transport. Let us write $\overline{%
\mathcal{E}}^{\overline{B}}\equiv \overline{\mathbf{\mathcal{E}}}\cdot 
\overline{\mathbf{B}}$\footnote[3]{Observe that in contrast to previous works, here 
$\overline{\mathcal{E}}^{\overline{B}}$ is the dot product of the t.e.m.f.
with $\overline{\mathbf{B}}$, and not the component of that force along the
mean magnetic field. We chose to work with this function because it 
facilitates the numerical integration of the equations, while preserving
the generic behaviour of the projections. }. 
Its evolution equation is $\partial \overline{%
\mathcal{E}}^{\overline{B}}/\partial t=\left( \partial \overline{\mathbf{
\mathcal{E}}}/\partial t\right) \cdot \overline{\mathbf{B}}+\overline{
\mathbf{\mathcal{E}}}\cdot \partial \overline{\mathbf{B}}/\partial t$.
Proceeding similarly as in Refs. \citep{bf2002,yo}, we obtain the following
full form for the evolution equation for $\overline{\mathcal{E}}^{\overline{B
}}$, in the two scale approximation: 
\begin{eqnarray}
\frac{\partial \overline{\mathcal{E}}^{\overline{B}}}{\partial t}&=&\frac{2}{3}
k_{L}H_{S}^{C}H_{L}^{C}+\frac{1}{3}k_{L}\left[ k_{S}^{2}H_{S}^{M}-H^{u}
\right] \left\vert H_{L}^{M}\right\vert \nonumber\\
&-&\frac{2}{3}k_{L}^{2}E^{u}H_{L}^{M}
-\zeta _{1}\overline{\mathcal{E}}^{\overline{B}},  \label{es-a}
\end{eqnarray}
where we replaced $\overline{\overline{\mathbf{U}}\cdot \overline{\mathbf{B}}%
}\simeq \left\langle \overline{\mathbf{U}}\cdot \overline{\mathbf{B}}%
\right\rangle _{vol}=H_{L}^{C}$, the large scale cross-helicity; 
$H_{S}^{C}=\left\langle \mathbf{u}\cdot \mathbf{b}\right\rangle _{vol}\simeq 
\overline{\mathbf{u}\cdot \mathbf{b}}$, the small scale cross-helicity; $%
\overline{\left( \mathbf{\nabla }\times \mathbf{b}\right) \cdot \mathbf{b}}%
\simeq k_{S}^{2}H_{S}^{M}$; $E^{u}=\overline{u^{2}}/2$ and $H^{u}=\overline{
\mathbf{u}\cdot \left( \mathbf{\nabla }\times \mathbf{u}\right) }$ where these
two last quantities are considered prescribed. Since we are considering $\overline{%
\mathbf{B}}$ to be force-free, we replaced $\left\vert \overline{\mathbf{B}}%
\right\vert ^{2}\simeq k_{L}\left\vert H_{L}^{M}\right\vert $ and $\left( 
\mathbf{\nabla }\times \overline{\mathbf{B}}\right) \cdot \overline{\mathbf{B%
}}\simeq k_{L}^{2}H_{L}^{M}$. The last term in eq. (\ref{es-a}), $\zeta _{1}
\overline{\mathcal{E}}^{B}$, includes the effect of viscosity, resistivity, 
the term $\overline{\mathbf{\mathcal{E}}}\cdot \partial \overline{\mathbf{B}}/
\partial t$ and,
more importantly, the three point correlations denoted by $\overline{\mathbf{%
T}}$ in eq. (\ref{r}). We see that besides the equations derived until
now we need the ones for $H_{L,S}^{C}$. The derivation of these equations is
sketched in the Appendix and, although being a straightforward procedure, it
is a rather tedious one and we end up with a system of seven equations,
besides the four ones already shown above: the two equations for $%
H_{L,S}^{C} $, the ones for $\overline{F}^{\overline{B}}\equiv \overline{%
\left( \mathbf{\nabla }\times \mathbf{b}\right) \times \mathbf{b}}\cdot 
\overline{\mathbf{B}}$, i.e., the dot product of the small scale Lorentz
force with $\overline{\mathbf{B}}$, for $\overline{F}^{\overline{W}}\equiv 
\overline{\left( \mathbf{\nabla }\times \mathbf{b}\right) \times \mathbf{b}}%
\cdot \overline{\mathbf{W}}$, for $\overline{F}^{\overline{U}}\equiv 
\overline{\left( \mathbf{\nabla }\times \mathbf{b}\right) \times \mathbf{b}}%
\cdot \overline{\mathbf{U}}$, for $\overline{\mathcal{E}}^{W}=\overline{%
\mathbf{\mathcal{E}}}\cdot \overline{\mathbf{W}}$, i.e., the scalar product
of the t.e.m.f. with the large scale vorticity, and for $E^{b}=\overline{%
b^{2}}/2$, i.e., the small scale magnetic energy. Eq. (\ref{es-a}) is
already expressed in the two scale form. The other ten equations read 
\begin{equation}
\frac{\partial H_{L}^{M}}{\partial t}\simeq 2\overline{\mathcal{E}}^{%
\overline{B}}-2\eta k_{L}^{2}H_{L}^{M},  \label{es-b}
\end{equation}%
\begin{equation}
\frac{\partial H_{S}^{M}}{\partial t}\simeq -2\overline{\mathcal{E}}^{%
\overline{B}}-2\eta k_{S}^{2}H_{S}^{M},  \label{es-aa}
\end{equation}%
\begin{equation}
\frac{\partial H^{\overline{U}}}{\partial t}\simeq 2\overline{F}^{\overline{W%
}}-2\nu k_{L}^{2}H^{\overline{U}},  \label{es-m}
\end{equation}%
\begin{eqnarray}
\frac{\partial \overline{\mathcal{E}}^{\overline{W}}}{\partial t}&=&\frac{2}{3}%
k_{L}H_{S}^{C}\left\vert H^{U_{0}}\right\vert +\frac{1}{3}k_{L}\left[
k_{S}^{2}H_{S}^{M}-H^{u}\right. \nonumber\\
&&\left. -2k_{L}E^{u}\right] H_{L}^{C} 
-\zeta_{2}\overline{\mathcal{E}}^{\overline{W}},  \label{es-d}
\end{eqnarray}
\begin{equation}
\frac{\partial H_{L}^{C}}{\partial t}=\overline{F}^{\overline{B}}+\overline{%
\mathcal{E}}^{\overline{W}}-2\left( \nu +\eta \right) k_{L}^{2}H_{L}^{C},
\label{es-g}
\end{equation}%
\begin{equation}
\frac{\partial H_{S}^{C}}{\partial t}=-\overline{F}^{\overline{B}}-\overline{%
\mathcal{E}}^{\overline{W}}-2\left( \nu +\eta \right) H_{L}^{C},
\label{es-h}
\end{equation}%
\begin{equation}
\frac{\partial \overline{F}^{\overline{W}}}{\partial t}=\frac{1}{3}%
k_{L}^{3}H_{S}^{C}H_{L}^{C}-\frac{2}{3}k_{L}^{2}\sqrt{2E_{2}^{b}}H^{%
\overline{U}}-\zeta _{F}\overline{F}^{\overline{W}},  \label{es-i}
\end{equation}%
\begin{equation}
\frac{\partial \overline{F}^{\overline{U}}}{\partial t}\simeq \frac{1}{3}%
k_{L}^{2}H_{S}^{C}H_{L}^{C}-\frac{2}{3}k_{L}\sqrt{2E_{2}^{b}}\left\vert H^{%
\overline{U}}\right\vert -\zeta _{F}\overline{F}^{\overline{U}},
\label{es-j}
\end{equation}%
\begin{equation}
\frac{\partial \overline{F}^{\overline{B}}}{\partial t}\simeq \frac{1}{3}%
k_{L}^{3}H_{S}^{C}\left\vert H_{L}^{M}\right\vert -\frac{2}{3}k_{L}^{2}\sqrt{%
2E_{2}^{b}}H_{L}^{C}-\zeta _{F}\overline{F}^{\overline{B}},  \label{es-k}
\end{equation}%
and 
\begin{equation}
\frac{\partial E_{2}^{b}}{\partial t}=-2\overline{F}^{\overline{U}}\sqrt{%
E_{2}^{b}}-k_{S}^{2}H_{S}^{M}\overline{\mathcal{E}}^{\overline{B}}+H_{S}^{C}%
\overline{F}^{\overline{B}}-\zeta _{b}E_{2}^{b},  \label{es-l}
\end{equation}%
where $E_{2}^{b}=\left( E^{b}\right) ^{2}$.

\section{Making the Equations Non-Dimensional}

In order to work with non-dimensional quantities, we define the following
dimensionless variables: $\tau =k_{S}ut$, $\mathcal{F}^{\mathcal{W}}=
\overline{F}^{\overline{W}}/k_{S}^{2}u^{3}$, $\mathcal{F}^{\mathcal{U}}=
\overline{F}^{\overline{U}}/k_{S}u^{3}$, $\mathcal{F}^{\mathcal{B}}=
\overline{F}^{\overline{B}}/k_{S}u^{3}$, $\Xi _{2}^{b}=E_{2}^{b}/u^{4}$, 
$\mathcal{Q}^{\mathcal{B}}=\overline{\mathcal{E}}^{\overline{B}}/u^{3}$, 
$\mathcal{Q}^{\mathcal{W}}=\overline{\mathcal{E}}^{\overline{W}}/k_{S}u^{3}$, 
$\mathcal{H}_{S,L}^{M}=k_{S}H_{S,L}^{M}/u^{2}$, $\mathcal{H}^{\overline{U}
}=H^{\overline{U}}/k_{S}u^{2}$, $\mathcal{H}_{L,S}^{C}=H_{L,S}^{C}/u^{2}$, 
$\mathcal{H}^{u}=H^{u}/k_{S}u^{2}$, $\Xi ^{u}=E^{u}u^{2}$, $\xi _{i}=\zeta
_{i}/k_{S}u$ ($i=1,2,F,b$), $R_{e}=\nu k_{S}/u$, $R_{m}=\eta k_{S}/u$, with 
$R_{e}$ the Reynolds number, $R_{m}$ the magnetic Reynolds number, and 
$r=k_{L}/k_{S}$. The non-dimensional equations then read 
\begin{eqnarray}
\frac{\partial \mathcal{Q}^{\mathcal{B}}}{\partial \tau }&=&\frac{2}{3}r
\mathcal{H}_{S}^{C}\mathcal{H}_{L}^{C}+\frac{1}{3}r\left[ \mathcal{H}_{S}^{M}
-\mathcal{H}^{u}\right] \left\vert \mathcal{H}_{L}^{M}\right\vert \nonumber\\
&-&\frac{2}{3}r^{2}\Xi ^{u}\mathcal{H}_{L}^{M}-\xi _{1}\mathcal{Q}^{\mathcal{B}
},  \label{eqadi-1}
\end{eqnarray}
\begin{eqnarray}
\frac{\partial \mathcal{Q}^{\mathcal{W}}}{\partial \tau }&=&\frac{2}{3}r%
\mathcal{H}_{S}^{C}\left\vert \mathcal{H}^{\overline{U}}\right\vert +\frac{1%
}{3}r\left[ \mathcal{H}_{S}^{M}-\mathcal{H}^{u} -2r\Xi ^{u}\right] 
\mathcal{H}_{L}^{C} \nonumber\\
&-&\xi _{2}\mathcal{Q}^{\mathcal{W}},  
\label{eqadi-2}
\end{eqnarray}
\begin{equation}
\frac{\partial \mathcal{H}_{L}^{M}}{\partial \tau }=2\mathcal{Q}^{\mathcal{B}%
}-\frac{2r^{2}}{R_{m}}\mathcal{H}_{L}^{M},  \label{eqadi-3}
\end{equation}%
\begin{equation}
\frac{\partial \mathcal{H}_{S}^{M}}{\partial \tau }=-2\mathcal{Q}^{\mathcal{B%
}}-\frac{2}{R_{m}}\mathcal{H}_{S}^{M},  \label{eqadi-4}
\end{equation}%
\begin{equation}
\frac{\partial \mathcal{H}^{\overline{U}}}{\partial t}\simeq 2\mathcal{F}^{%
\mathcal{W}}-\frac{2r^{3}}{R_{e}}\mathcal{H}^{\overline{U}},  \label{eqadi-5}
\end{equation}%
\begin{equation}
\frac{\partial \mathcal{H}_{L}^{C}}{\partial \tau }=\mathcal{F}^{\mathcal{B}%
}+\mathcal{Q}^{\mathcal{W}}-\left( \frac{2}{R_{m}}+\frac{2}{R_{e}}\right)
r^{2}\mathcal{H}_{L}^{C},  \label{eqadi-6}
\end{equation}%
\begin{equation}
\frac{\partial \mathcal{H}_{S}^{C}}{\partial \tau }=-\mathcal{F}^{\mathcal{B}%
}-\mathcal{Q}^{\mathcal{W}}-\left( \frac{2}{R_{m}}+\frac{2}{R_{e}}\right) 
\mathcal{H}_{S}^{C},  \label{eqadi-7}
\end{equation}%
\begin{equation}
\frac{\partial \mathcal{F}^{\mathcal{W}}}{\partial \tau }=\frac{1}{3}r^{3}%
\mathcal{H}_{S}^{C}\mathcal{H}_{L}^{C}-\frac{2}{3}r^{2}\sqrt{2\Xi _{2}^{b}}%
\mathcal{H}^{\overline{U}}-\xi _{F}\mathcal{F}^{\mathcal{W}},
\label{eqadi-8}
\end{equation}%
\begin{equation}
\frac{\partial \mathcal{F}^{\mathcal{U}}}{\partial \tau }\simeq \frac{1}{3}%
r^{2}\mathcal{H}_{S}^{C}\mathcal{H}_{L}^{C}-\frac{2}{3}r\sqrt{2\Xi _{2}^{b}}%
\left\vert \mathcal{H}^{\overline{U}}\right\vert -\xi _{F}\mathcal{F}^{%
\mathcal{U}},  \label{eqadi-9}
\end{equation}%
\begin{equation}
\frac{\partial \mathcal{F}^{\mathcal{B}}}{\partial \tau }\simeq \frac{1}{3}%
r^{3}\mathcal{H}_{S}^{C}\left\vert \mathcal{H}_{L}^{M}\right\vert -\frac{2}{3%
}r^{2}\sqrt{2\Xi _{2}^{b}}\mathcal{H}_{L}^{C}-\xi _{F}\mathcal{F}^{\mathcal{B%
}},  \label{eqadi-10}
\end{equation}%
and 
\begin{equation}
\frac{\partial \Xi _{2}^{b}}{\partial \tau }\simeq -2\mathcal{F}^{\mathcal{U}%
}\sqrt{\Xi _{2}^{b}}-\mathcal{H}_{S}^{M}\mathcal{Q}^{\mathcal{B}}+\mathcal{H}%
_{S}^{C}\mathcal{F}^{\mathcal{B}}-\xi _{b}\Xi _{2}^{b}.  \label{eqadi-11}
\end{equation}

\section{Numerical Results and Discussion}

\begin{figure}
\includegraphics[width=75mm]{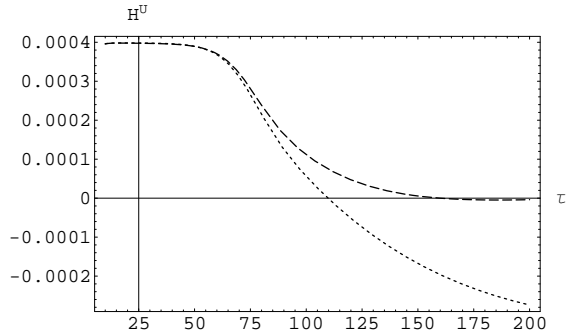} \vspace{0.5cm}
\caption{Large scale kinetic helicity $\mathcal{H}^{u}$ as a function of 
$\tau $ for $\xi_i =1/2$. The long dashed line corresponds to $R_m = 200$,
and the dotted line to $R_m=2000$. The generation of $H^{\overline U}$ is
stronger for the largest value of $R_m$.}
\label{huz1}
\end{figure}

\begin{figure}
\includegraphics[width=75mm]{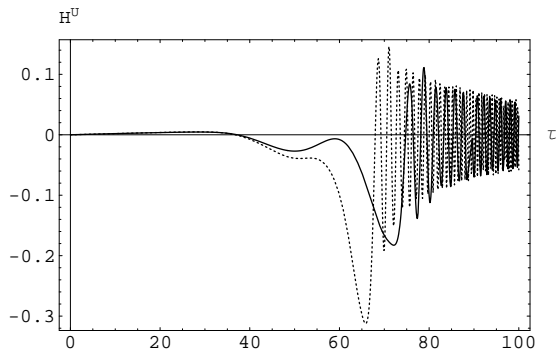} \vspace{0.5cm}
\caption{Large scale kinetic helicity $\mathcal{H}^{u}$ as a function of 
$\tau $ for $\xi_i=1/R_m$. The continuum line corresponds to $R_m=200$, and 
the dotted one to $R_m=200$. At the beginning, the induction of $H^{\overline U}$ is
stronger for the largest value of $R_m$.}
\label{huzRM}
\end{figure}

\begin{figure}
\includegraphics[width=75mm]{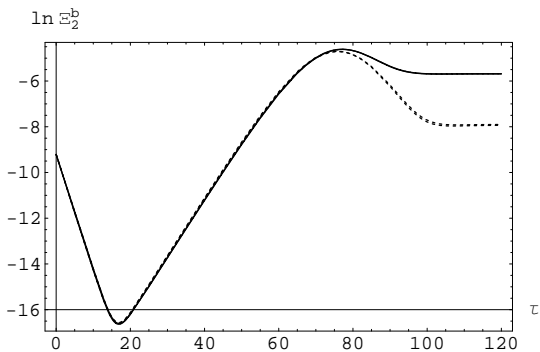} \vspace{1cm}
\caption{Logarith of the small scale magnetic energy $\Xi^b_2$ as a
function of $\tau$, for $\xi_i=1/2$. Each curve is in fact two curves, with
and without the effect of $\overline{U}$, which means that in this case
the effect of those flows is negligible. The upper curve corresponds to
$R_m=2000$ and the lower one to $R_m=200$. The small scale magnetic energy 
density is very small although larger for $R_m=2000$}
\label{ebz1}
\end{figure}

\begin{figure}
\includegraphics[width=75mm]{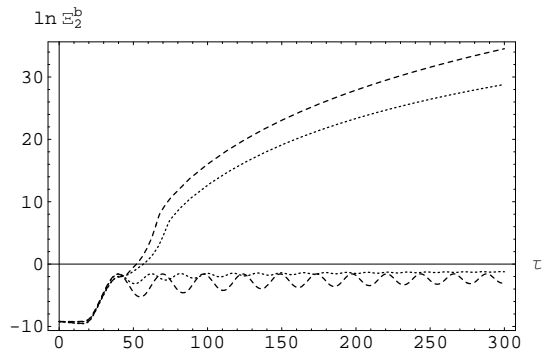} \vspace{1cm}
\caption{Logarith of the small scale magnetic energy $\Xi^b_2$ as a
function of $\tau$ for $\xi_i=1/R_m$. The upper, long dash curve corresponds 
to the presence of $\overline{U}$, while the lower, oscillating one, to
the absence of those flows, both of them for $R_m=2000$. Short dashed
curves represents the same quantities but for $R_m=200$. In this case
large scale flows strongly enhance the production of small scale magnetic
energy, and this effect is again stronger for larger values of $R_m$.}
\label{ebzRM}
\end{figure}

\begin{figure}
\includegraphics[width=75mm]{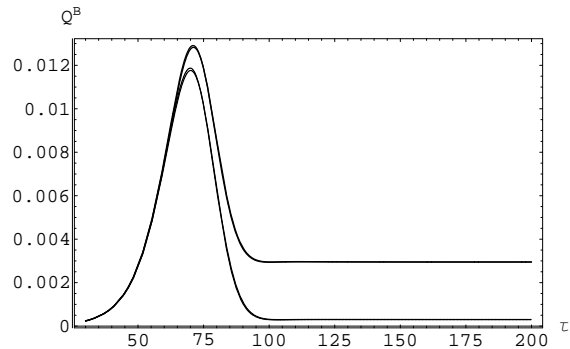} \vspace{1cm}
\caption{Mean electromotive force $\mathcal{Q}^{\cal B}$ as function of 
$\tau$ for $\xi_i=1/2$. Each curve is two curves, one with the effect of 
$\overline{U}$ and the other without those fields, showing that for this 
value of $\xi$ the effect of those fields is negligible. Upper curve 
corresponds to $R_m=2000$ and lower curve to $R_m=200$, which shows 
that for the largest $R_m$, $\mathcal{Q}^{\cal B}$ is slightly stronger . }
\label{qbz1}
\end{figure}

\begin{figure}
\includegraphics[width=75mm]{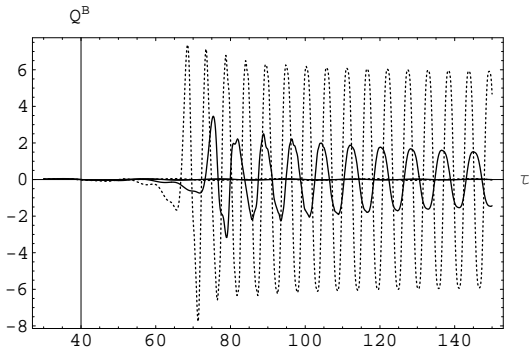} \vspace{1cm}
\caption{Mean electromotive force $\mathcal{Q}^{\cal B}$ as function of 
$\tau$ for $\xi_i=1/R_m$. Dotted line corresponds to $R_m=2000$ and
continuum line to $R_m=200$. The curves corresponding to the absence
of large scale flows are of negligible amplitude and almost indistinguishable
from the $\tau$ axis, showing that in this case the enhancement of the
t.e.m.f. by the shear fields is very strong. }
\label{qbzRM}
\end{figure}

\begin{figure}
\includegraphics[width=75mm]{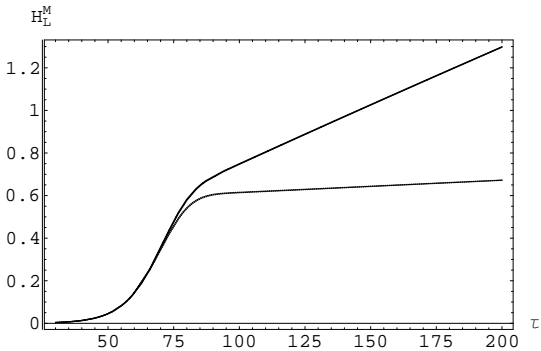} \vspace{1cm}
\caption{Large scale magnetic helicity $\mathcal{H}^M_L$ as a function of
$\tau$, for $\xi_i=1/2$. Again in this figure, each curve is two curves,
one with the effect of $\overline{U}$ and the other without, showing 
that the effect of those flows on the evolution of magnetic helicity
is negligible for the chosen value of $\xi_i$. The growing curve correspond
to $R_m=200$, while the slowly growing, lower curve to $R_m=2000$.}
\label{hmlz1}
\end{figure}

\begin{figure}
\includegraphics[width=75mm]{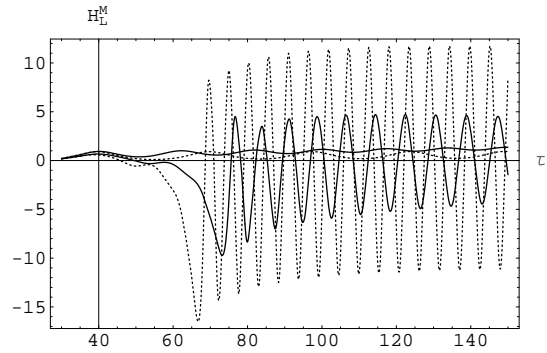} \vspace{1cm}
\caption{Large scale magnetic helicity $\mathcal{H}^M_L$ as a function of
$\tau$, for $\xi_i=1/R_m$. Dotted curves correspond to $R_m=2000$ while
continuous curves to $R_m=200$. In each case, the strongly oscillating
curves correspond to the action of large scale flows, while the slowly
oscillations to their absence. Consistently with what was shown in 
Fig. \ref{qbzRM}, the action of $\overline{U}$ enhances the cascade of
magnetic helicity.}
\label{hmlzRM}
\end{figure}

\begin{figure}
\includegraphics[width=75mm]{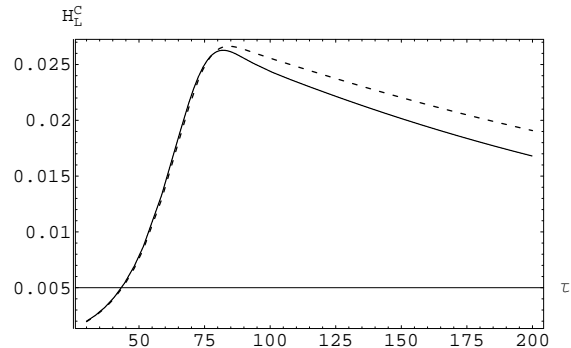} \vspace{1cm}
\caption{Large scale cross helicity $\mathcal{H}^C_L$ as function of 
$\tau$ for $\xi_i=1/2$. Consistently with Fig. \ref{huz1} the generation
of large scale cross-helicity is stronger for $R_m=2000$ (dashed line)
than for $R_m=200$ (full line).}
\label{hclz1}
\end{figure}

\begin{figure}
\includegraphics[width=75mm]{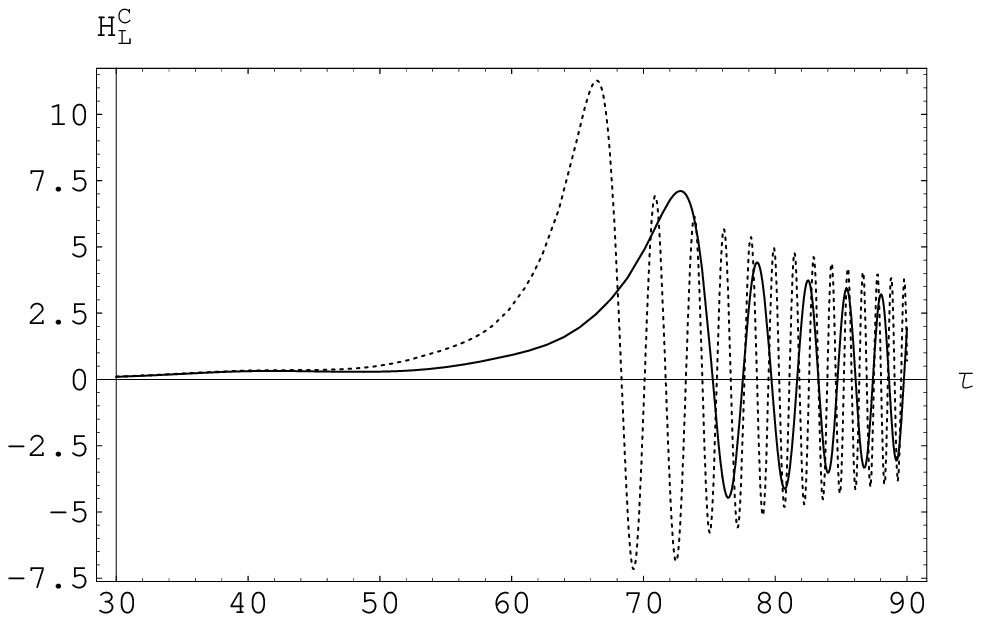} \vspace{1cm}
\caption{Large scale cross helicity $\mathcal{H}^C_L$ as function of 
$\protect\tau$, for $\xi_i=1/R_m$. Consistently with Fig. 
\ref{huzRM} the generation of large scale cross-helicity is stronger
for $R_m=2000$ (dotted line) than for $R_m=200$ (continuous line).}
\label{hclzRM}
\end{figure}

We numerically integrated equations (\ref{eqadi-1})-(\ref{eqadi-11}) using
the following parameters and initial conditions: $r=0.2$, $\mathcal{H}^{u}=-1
$, $\Xi ^{u}=1$,  $\mathcal{Q}^{%
\mathcal{B}}\left( 0\right) =\mathcal{Q}^{\mathcal{W}}\left( 0\right) =0$, $%
\mathcal{H}_{L}^{M}\left( 0\right) =0.001$, $\mathcal{H}_{0}^{M}\left(
0\right) =-0.001$, $\mathcal{H}^{\overline{U}}\left( 0\right) =0.0001$, $%
\mathcal{H}_{L}^{C}\left( 0\right) =0.0001$, $\mathcal{H}_{S}^{C}\left(
0\right) =-0.0001$, $\mathcal{F}^{\mathcal{B}}\left( 0\right) =\mathcal{F}^{%
\mathcal{W}}\left( 0\right) =\mathcal{F}^{\mathcal{U}}\left( 0\right) =0$, $%
\Xi _{2}^{b}\left( 0\right) =0$, $R_{m}=R_{e}=200$ and $2000$ (i.e., magnetic 
Prandtl number $P_{m}=1$), $\xi _i=1/2$ (strong non-linearities) and 
$\xi _i =1/R_{m}$ (weak non-linearities)\footnote[4]{As before, subindex
 "$i$ " denotes $1$, $2$, $F$ and $b$. In each integration we assume that 
all $\xi_i$ are the same for all ``$i$''.}.
A comment about the chosen values for $\xi_i$ is in order: In principle this
parameter can depend on $R_m$; however, results of numerical simulations show
that it is of order unity for $R_m < 100$. In this sense, the value $\xi=1/2$
would be in accord with those results. As we are working here with larger 
values of $R_m$ for which, to our knowledge, there lacks numerical estimations
of $\xi_i$, we chose two values that might represent the two extreme behaviours
of this parameter. Nevertheless, we must stress that the validity of this choice
should be checked by direct numerical simulations.
In Fig. \ref{huz1} we plotted $\mathcal{H}^{\overline{U}}$ as a function of 
$\tau $ for $\xi _i=1/2$. The long dashed line corresponds to $R_{m}=200$ 
while the short dashed one to $R_{m}=2000$. We see that the generation of  
$\overline{\mathbf{U}}$ is rather weak, but the effect seems to be stronger
for $R_m=2000$ as time passes.
In Fig. \ref{huzRM} we plotted $\mathcal{H}^{\overline{U}}$ as a function
of $\tau $ for $\xi_i=1/R_{m}$. The full line corresponds to $R_{m}=200$ while
the doted one to $R_{m}=2000$. In this case there is a strong production of
large scale kinetic helicity, it being stronger for $R_{m}=2000$
at the beginning of the integration, while for later times there seems to
be no difference between the outcomes for the two $R_m$ considered.
In Fig. \ref{ebz1} we plotted the the logarithm of the small scale
magnetic energy,  $ln\left( \Xi ^{b}_2\right) $ as a function of $\tau $, 
for $\xi_i=1/2$. Each curve consists of two curves: one with the effect of 
$\overline{\mathbf{U}}$ and the other without this field. This superposition
of curves means that for the chosen value of $\xi _i$ the effect of 
$\overline{\mathbf{U}}$ on the evolution of small scale magnetic energy is 
negligible. The upper curve corresponds to the largest value of $R_{m}$, 
and we see that in this case a
saturation value for $ \Xi ^{b}_2$ larger than for $R_m=200$ is attained.
In Fig. \ref{ebzRM} we plotted the logarithm of the small scale magnetic
energy $ln\left( \Xi _{b}\right) $, for $\xi _i=1/R_{m}$. Long dash curves
correspond to $R_{m}=2000$: upper curve contains the effect of $\overline{\bf U}$, 
lower oscillating curve is without the action of those fields. Short dashed curves
correspond to $R_m=200$, with the same features for the presence and absence
of $\overline{\bf U}$. We see that the action of $\overline{\bf U}$ strongly
enhances the generation of small scale magnetic energy, and again this 
effect is stronger for larger $R_m$.
In Fig. \ref{qbz1} we plotted $\mathcal{Q}^{\mathcal{B}}$ as a function of 
$\tau $ for $\xi_i=1/2$. Here again each curve consists of two curves, one
with the effect of $\overline{\mathbf{U}}$ and the other without, showing
again that for strong non-linearities the effect of those flows is
negligible, consistently with previous figures. Upper curve corresponds 
to $R_{m}=2000$ while lower curve to $R_{m}=200$. We see that again for 
larger $R_{m}$ there is an enhancement of $\mathcal{Q}^{\mathcal{B}}$.
In Fig. \ref{qbzRM} we plotted $\mathcal{Q}^{\mathcal{B}}$ as a function of 
$\tau $ for $\xi_i =1/R_{m}$. We see here again that the action of large
scale flows enhances the mean electromotive force $\mathcal{Q}^{\mathcal{B}}$
and this enhancement is stronger for larger $R_{m}$. Oscillating, 
dotted-line curve corresponds to $R_{m}=2000$, and full-line curve to 
$R_m=200$. The curves corresponding to the absence of the effect of those 
flows are almost indistinguishable from the $\tau$ axis. 
In Fig. \ref{hmlz1} we plotted $\mathcal{H}_{L}^{M}$ as a function of $\tau 
$ for $\xi =1/2$. Here again each curve consists of two curves, one with
the effect of $\overline{\mathbf{U}}$ and the other without, showing again
that for strong non-linearities the effect of those flows is negligible.
Fast growing curve corresponds to $R_{m}=200$ while lower one to $R_{m}=2000$. 
 The coincidence of the two curves for short times
corresponds to the kinematic regime, where back-reaction of the induced
magnetic fields $\mathbf{b}$ did not take place yet.
In Fig. \ref{hmlzRM} we plotted $\mathcal{H}_{L}^{M}$ as a function of $\tau 
$ for $\xi =1/R_{m}$. We see here again that the action of large scale
flows enhances the mean electromotive force $\mathcal{H}_{L}^{M}$ and this
enhancement is stronger for larger $R_{m}$. Dashed curves correspond to 
$R_{m}=2000$: the ones with the largest amplitude correspond to 
the action of $\overline{\mathbf{U}}$, while the lower amplitude to the absence 
of this effect. Full line correspond to $R_{m}=200$, and the features 
with respect to the presence and absence of large scale flows are the same as 
for $R_{m}=200$. The coincidence of all four curves at the beginning of the
evolution corresponds to the kinematic regime.
In Fig. \ref{hclz1} we plotted $\mathcal{H}_{L}^{C}$ as a function of $\tau 
$ for $\xi _i=1/2$. Dashed line curve corresponds to $R_{m}=2000$ while
full line to $R_{m}=200$. Consistently with Fig. (\ref{huz1}), we see
that $\mathcal{H}_{L}^{C}$ is larger for larger $R_{m}$.
In Fig. \ref{hclzRM} we plotted $\mathcal{H}_{L}^{C}$ as a function of 
$\tau $ for $\xi _i=1/R_{m}$. Dotted line corresponds to $R_{m}=2000$
while full line to $R_{m}=200$. Consistently with Fig. (\ref{huzRM}),
we see that $\mathcal{H}_{L}^{C}$ is larger for  $R_{m}=2000$ than for $R_m=200$,
with the difference in amplitudes between both quantities getting smaller with
time.

\section{Conclusions}

In this paper we studied semi-analytically and qualitatively the generation
of large scale flows by the action of a turbulent mean field dynamo, and the
back-reaction of those flows on the turbulent electromotive force for two values
of magnetic Reynolds number, $R_{m}=200$ and $2000$, and magnetic Prandtl number 
$P_{m}=1$. We considered a system in which small scale turbulent flows are
fully helical and prescribed by a given external mechanism, i.e., a kinematically
driven dynamo, and that this
system possesses boundary conditions such that all total divergencies
vanish. The turbulence was considered to be homogeneous and isotropic, which
although being of limited applicability to obtain quantitative results for real
systems, it serves to study many conceptual aspects of large scale magnetic
field generation, besides enormously simplifying the mathematics. We
followed the evolution of large scale flows through their associated kinetic
helicity $H^{\overline{U}}$, as one of the suppositions we made was that
those flows were fully helical too. One crucial assumption we made was that 
$\mathbf{\nabla }\cdot \overline{\mathbf{\mathcal{E}}}=0$, which
due to the Coulomb gauge, imposed a further constraint on $\overline{\mathbf{%
U}}$ and $\overline{\mathbf{B}}$ that allowed us to expressed several terms
as large scale cross-helicity. Another one was to assume that $\overline{%
\mathbf{B}}$ is force-free and the main reason to adopt it is that it helped
to simplify the mathematics. Although this condition can be fulfilled in
certain astrophysical environments, this is not a general situation,
therefore it should be dropped off in future works that aim at generalizing
this work.

We found that large scale flows act on the t.e.m.f. $\overline{\mathbf{%
\mathcal{E}}}$ through large and small scale cross-helicities, and that, for
the minimal $\tau $ closure considered here, the effect of those fields is
stronger for large relaxation times ( $\xi_i=1/R_m $). For short
relaxation time ($\xi_i=1/2$), the effect of those fields seems to be negligible. 
The choice of the values for $\xi_i$ was arbitrary, in the sense that, to
our present knowledge, it is not known how that parameter depends on the
magnetic Reynolds number for $R_m > 100$. For $R_m < 100$ it seems to be 
confirmed that $\xi_i$ is of order unity. Here we chose to work with two
values that may be considered as representative of two extreme possibilities:
$\xi_i=1/2$ would be consistent with the predictions of the numerical 
simulations (although they were made for a different $R_m$ interval), while
$\xi=1/R_m$ would represent a resistive case. In any case, more reliable values
should be given by numerical simulations performed for $R_m> 100$.

Due to the simple system considered and the approximations we made, we do
not intended to find quantitative results, as for example estimate the time
interval during which $\overline{\mathbf{U}}\times \overline{\mathbf{B}}\ll 
\overline{\mathbf{\mathcal{E}}}$ is valid, nor do we extract more conceptual
and qualitative conclusions. We end this work stressing the importance of
studying this problem via numerical simulations, that will show us the next
paths to follow in a further analytical study, besides confirming or
contesting the results presented here. The semi-analytical study of the
anisotropic case is also of the most importance, as well as the
consideration of other boundary conditions.

\section*{Acknowledgments}

We thank P. Mininni for suggesting the idea of this work, and the referee,
for many suggestions and comments that helped to improve it.
Financial support from FAPESB, grant APR205/2005 is acknowledge.
The present work was developed under project PROPP-UESC 00220.1300.489.

\appendix

\section{Deduction of the complementary evolution equations}

Here we sketch the derivation of the evolution equation for the large scale
kinetic helicity as well as the set of extra equations needed to study the
problem considered in this article.

\subsection{Evolution Equation for the Large Scale Vorticy}

We start from Navier-Stokes equation written in the form

\begin{eqnarray}
\frac{\partial \mathbf{U}}{\partial t}&=&-\left( \mathbf{\nabla }\times 
\mathbf{U}\right) \times \mathbf{U}-\mathbf{\nabla }\left( \frac{U^{2}}{2}+%
\frac{p}{\rho }\right) \nonumber\\
&+&\left( \mathbf{\nabla }\times \mathbf{B}\right)
\times \mathbf{B}+\nu \nabla ^{2}\mathbf{U}.  \label{apx-a}
\end{eqnarray}
The equation for $\overline{\mathbf{W}}=\mathbf{\nabla }\times \overline{%
\mathbf{U}}$ is obtained by simply taking the curl of eq. (\ref{apx-a}),
after replacing the decomposition in mean and stochastic fields, and using
the hypothesis that $\mathbf{u}$ is fully helical. We have

\begin{eqnarray}
\frac{\partial \overline{\mathbf{W}}}{\partial t}&=&-\mathbf{\nabla }\times
\left( \overline{\mathbf{W}}\times \overline{\mathbf{U}}\right) +\mathbf{%
\nabla }\times \overline{\left( \mathbf{\nabla }\times \mathbf{b}\right)
\times \mathbf{b}} \nonumber\\
&+&\nu \nabla ^{2}\overline{\mathbf{W}}.  \label{apx-b}
\end{eqnarray}

\subsection{Evolution Equation for the Large Scale Kinetic Helicity}

It is obtained as $\partial H^{\overline{U}}/\partial t=\left\langle 
\overline{\mathbf{U}}\cdot \partial \overline{\mathbf{W}}/\partial
t\right\rangle _{vol}+\left\langle \overline{\mathbf{W}}\cdot \partial 
\overline{\mathbf{U}}/\partial t\right\rangle _{vol}$. Replacing the
corresponding equations we obtain

\begin{equation}
\frac{\partial H^{\overline{U}}}{\partial t}\simeq 2\overline{\left( \mathbf{%
\nabla }\times \mathbf{b}\right) \times \mathbf{b}}\cdot \overline{\mathbf{W}%
}-2\nu k_{L}^{2}H^{\overline{U}},  \label{apx-c}
\end{equation}%
where the semi-equality stems from the fact that we approximated $%
\left\langle \cdots \right\rangle _{vol}\simeq \overline{\cdots }$. We
define $\overline{F}^{\overline{W}}\equiv \overline{\left( \mathbf{\nabla }%
\times \mathbf{b}\right) \times \mathbf{b}}\cdot \overline{\mathbf{W}}$, and
thus eq. (\ref{apx-c}) reads

\begin{equation}
\frac{\partial H^{\overline{U}}}{\partial t}=2\overline{F}^{\overline{W}%
}-2\nu k_{L}^{2}H^{\overline{U}}.  \label{apx-d}
\end{equation}

\subsection{Evolution Equation for $\overline{\mathbf{F}}\equiv \overline{%
\left( \protect\nabla \times \mathbf{b}\right) \times \mathbf{b}}$ and its
Projections}

It is found by taking curl of eq. (\ref{j}) and using it to expand $%
\overline{\left[ \partial \left( \mathbf{\nabla }\times \mathbf{b}\right)
/\partial t\right] \times \mathbf{b}}+\overline{\left( \mathbf{\nabla }%
\times \mathbf{b}\right) \times \partial \mathbf{b}/\partial t}$. After a
somewhat lengthy, but straightforward calculation, where it was assumed that
for the two scale approximation $\nabla \cdot \overline{\mathbf{F}}
\simeq 0$\footnote[5]{$\nabla \cdot \left\langle \left( \nabla \times b\right) 
\times
b\right\rangle =-b\cdot \nabla ^{2}b-\left\vert \nabla \times b\right\vert
^{2}\simeq k_{S}^{2}\left\vert b\right\vert ^{2}-k_{S}^{2}\left\vert
b\right\vert ^{2}=0$}, we obtain

\begin{eqnarray}
\frac{\partial \overline{\mathbf{F}}}{\partial t} &\simeq &\frac{1}{3}\overline{
\left( \nabla \cdot \overline{\mathbf{\mathcal{E}}}\right) \left( \nabla
\times \overline{\mathbf{B}}\right) }-\frac{k_{S}}{3}\overline{\left( \nabla
\cdot \overline{\mathbf{\mathcal{E}}}\right) \overline{\mathbf{B}}}
\nonumber\\
&-&\frac{1}{3}H_{S}^{C}\nabla ^{2}\overline{\mathbf{B}}
+\frac{2}{3}E^{b}\nabla ^{2}
\overline{\mathbf{U}}-\zeta _{F}\overline{\mathbf{F}}.  \label{apx-i}
\end{eqnarray}
As in the body of the paper, we assume that $\overline{\nabla \cdot \mathbf{%
\mathcal{E}}}=0$, so eq. (\ref{apx-i}) reduces to 
\begin{equation}
\frac{\partial \overline{\mathbf{F}}}{\partial t}\simeq -\frac{1}{3}%
H_{S}^{C}\nabla ^{2}\overline{\mathbf{B}}+\frac{2}{3}E^{b}\nabla ^{2}%
\overline{\mathbf{U}}-\zeta _{F}\overline{\mathbf{F}}.  \label{apx-j}
\end{equation}

To find the evolution equations for the scalar product of $\overline{\mathbf{%
F}}$ with $\overline{\mathbf{W}}$ and $\overline{\mathbf{B}}$, we use the
above defined expression $\overline{F}^{\overline{W}}$, and an analogous
expression for $\overline{\mathbf{B}}$. Again the evolution equation is
found by taking the time derivative of the complete expression. In the two
scale approximation we have%
\begin{eqnarray}
\frac{\partial \overline{F}^{\overline{W}}}{\partial t} &\simeq &-\frac{1}{3}%
H_{S}^{C}\nabla ^{2}\overline{\mathbf{B}}\cdot \overline{\mathbf{W}}+\frac{2%
}{3}E^{b}\nabla ^{2}\overline{\mathbf{U}}\cdot \overline{\mathbf{W}}-\zeta
_{F}\overline{F}^{\overline{W}}  \nonumber \\
&=&\frac{k_{L}^{2}}{3}H_{S}^{C}\overline{\mathbf{B}}\cdot \overline{\mathbf{W%
}}-\frac{2k_{L}^{2}}{3}E^{b}H^{\overline{U}}-\zeta _{F}\overline{F}^{%
\overline{W}}.  \label{apx-k}
\end{eqnarray}%
Due to the fact that $\mathbf{\nabla }\cdot \overline{\mathbf{\mathcal{E}}}%
=0\Rightarrow \nabla \cdot \left( \overline{\mathbf{U}}\times \overline{%
\mathbf{B}}\right) \simeq 0$, we can write $\overline{\mathbf{B}}\cdot 
\overline{\mathbf{W}}\simeq \overline{\mathbf{U}}\cdot \left( \mathbf{\nabla 
}\times \overline{\mathbf{B}}\right) \simeq k_{L}\overline{\mathbf{U}}\cdot 
\overline{\mathbf{B}}\simeq k_{L}H_{L}^{C}$. Using the fact that for a fully
helical $\overline{\mathbf{U}}$ field we can write $\left\vert \overline{W}%
\right\vert \simeq k_{L}^{1/2}\left\vert H^{\overline{U}}\right\vert ^{1/2}$%
, we obtain 
\begin{equation}
\frac{\partial \overline{F}^{\overline{W}}}{\partial t}\simeq \frac{k_{L}^{3}%
}{3}H_{S}^{C}H_{L}^{C}-\frac{2k_{L}^{2}}{3}E^{b}H^{\overline{U}}-\zeta _{F}%
\overline{F}^{\overline{W}}.  \label{apx-l}
\end{equation}%
For the projection of $\overline{\mathbf{F}}$ along $\overline{\mathbf{B}}$
and along $\overline{\mathbf{U}}$ we proceed analogously as for $\overline{F}%
^{W}$. Using the fact that for a large scale force-free field we can write $%
\left\vert \mathbf{B}_{0}\right\vert =k_{L}^{1/2}\left\vert
H_{L}^{M}\right\vert ^{1/2}$, we obtain 
\begin{equation}
\frac{\partial \overline{F}^{\overline{B}}}{\partial t}\simeq \frac{k_{L}^{3}%
}{3}H_{S}^{C}\left\vert H_{L}^{M}\right\vert -\frac{2k_{L}^{2}}{3}%
E^{b}H_{L}^{C}-\zeta _{F}\overline{F}^{\overline{B}},  \label{apx-m}
\end{equation}%
and 
\begin{equation}
\frac{\partial \overline{F}^{\overline{U}}}{\partial t}\simeq \frac{k_{L}^{2}%
}{3}H_{S}^{C}H_{L}^{C}-\frac{2k_{L}}{3}E^{b}\left\vert H^{\overline{U}%
}\right\vert -\zeta _{F}\overline{F}^{\overline{U}}  \label{apx-n}
\end{equation}%
where we used $\left\vert \overline{\mathbf{U}}\right\vert ^{2}\simeq
\left\vert H^{\overline{U}}\right\vert /k_{L}$.

\subsection{Evolution Equation for the Cross-Helicity}

Cross-Helicity is defined as $H^{C}=\langle \mathbf{U}\cdot \mathbf{B}%
\rangle _{vol}$. After obtaining from eq. (\ref{apx-a}) the evolution
equations for $\overline{\mathbf{U}}$ and $\mathbf{u}$ and using eq. (\ref{g}%
) and (\ref{j}), we obtain the following equation for the large scale cross
helicity, $H_{L}^{C}$ and the small scale cross helicity $H_{s}^{C}$: 
\begin{eqnarray}
\frac{\partial }{\partial t}H_{L}^{C} &\simeq &\overline{\left( \mathbf{\
\nabla }\times \mathbf{b}\right) \times \mathbf{b}}\cdot \overline{\mathbf{B}%
}+\overline{\mathbf{\mathcal{E}}}\cdot \overline{\mathbf{W}} \nonumber \\
&-&2\left( \nu +\eta \right) k_{L}^{2}H_{L}^{C}  \label{apx-e}
\end{eqnarray}
and 
\begin{eqnarray}
\frac{\partial }{\partial t}H_{S}^{C} &\simeq & -\overline{\left( \mathbf{\
\nabla }\times \mathbf{b}\right) \times \mathbf{b}}\cdot \overline{\mathbf{B}%
}-\overline{\mathbf{\mathcal{E}}}\cdot \overline{\mathbf{W}} \nonumber\\
&-& 2\left( \nu +\eta \right) k_{S}^{2}H_{S}^{C}.  \label{apx-f}
\end{eqnarray}
Replacing $\overline{F}^{\overline{B}}\equiv \overline{\left( \mathbf{\nabla 
}\times \mathbf{b}\right) \times \mathbf{b}}\cdot \overline{\mathbf{B}}$,
defining $\overline{\mathcal{E}}^{\overline{W}}\equiv \overline{\mathbf{%
\mathcal{E}}}\cdot \overline{\mathbf{W}}$, and using the fact that for fully
helical $\overline{\mathbf{U}}$ we can write $\left\vert \overline{\mathbf{W}%
}\right\vert \simeq k_{L}^{1/2}\left\vert H^{\overline{U}}\right\vert ^{1/2}$%
, in the two scale approximation we have 
\begin{equation}
\frac{\partial H_{L}^{C}}{\partial t}\simeq \overline{\mathcal{E}}^{%
\overline{W}}+\overline{F}^{\overline{B}}-2\left( \nu +\eta \right)
k_{L}^{2}H_{L}^{C}  \label{apx-g}
\end{equation}%
and 
\begin{equation}
\frac{\partial H_{S}^{C}}{\partial t}\simeq -\overline{\mathcal{E}}^{%
\overline{W}}-\overline{F}^{\overline{B}}-2\left( \nu +\eta \right)
k_{S}^{2}H_{S}^{C}.  \label{apx-h}
\end{equation}

\subsection{Evolution Equation for $\overline{\mathcal{E}}^{\overline{W}}=%
\overline{\mathbf{\mathcal{E}}}\cdot \overline{\mathbf{W}}$}

Using equation (\ref{r}), and $\mathbf{\nabla }\cdot \overline{\mathbf{%
\mathcal{E}}}=0$, we obtain

\begin{eqnarray}
\frac{\partial \overline{\mathcal{E}}^{\overline{W}}}{\partial t} &\simeq & 
\frac{2}{3}H_{S}^{C}\left\vert \overline{\mathbf{W}}\right\vert ^{2}+
\frac{k_{L}}{3}\left( k_{S}^{2}H_{S}^{M}-H^{u}\right) H_{L}^{C} \nonumber\\
&-&\frac{1}{3}\overline{u^{2}}\left( \mathbf{\nabla }\times 
\overline{\mathbf{B}}\right) \cdot \overline{\mathbf{W}}
-\zeta_{2}\overline{\mathcal{E}}^{\overline{W}},
\label{apx-o}
\end{eqnarray}
where in the last term we considered the term $\overline{\mathbf{\mathcal{E}}%
}\cdot \partial \overline{\mathbf{W}}/\partial t$, and the three point
correlations. Performing $\left( \mathbf{\nabla }\times \overline{\mathbf{B}}%
\right) \cdot \overline{\mathbf{W}}\simeq k_{L}\overline{\mathbf{B}}\cdot 
\overline{\mathbf{W}}\simeq k_{L}\left( \mathbf{\nabla }\times \overline{%
\mathbf{B}}\right) \cdot \overline{\mathbf{U}}\simeq k_{L}^{2}H_{L}^{C}$,
where the semi-equality before the last stems from the fact that $\mathbf{%
\nabla }\cdot \overline{\mathbf{\mathcal{E}}}=-\mathbf{\ \nabla }\cdot
\left( \overline{\mathbf{U}}\times \overline{\mathbf{B}}\right) \simeq 0$,
we obtain 
\begin{eqnarray}
\frac{\partial \overline{\mathcal{E}}^{\overline{W}}}{\partial t} &\simeq &
\frac{2}{3}k_{L}H_{S}^{C}\left\vert H^{U_{0}}\right\vert +\frac{1}{3}k_{L}\left[
k_{S}^{2}H_{S}^{M}-H^{u}-2k_{L}E^{u}\right] H_{L}^{C}\nonumber\\
&-&\zeta _{2}\overline{\mathcal{E}}^{\overline{W}}.  \label{apx-p}
\end{eqnarray}

\subsection{Evolution Equation for $E^{b}$}

It is obtained by scalar multiplying eq. (\ref{j}) by $\mathbf{b}$ and then
taking volume average. In order to simplify the mathematics, we approximate
the volume averages by a dot product between spatial averages of functions
of stochastic and mean fields. 
\begin{equation}
\frac{\partial E^{b}}{\partial t}\simeq -\overline{\left( \nabla \times 
\mathbf{b}\right) \times \mathbf{b}}\cdot \overline{\mathbf{U}}+\overline{%
\left( \nabla \times \mathbf{b}\right) \times \mathbf{u}}\cdot \overline{%
\mathbf{B}}-\zeta _{b}E^{b}.  \label{apx-q}
\end{equation}%
To deal with the second term we write $\nabla \times \mathbf{b}=\left[
\left( \nabla \times \mathbf{b}\right) \cdot \mathbf{b}\right] \mathbf{b}%
/\left\vert \mathbf{b}\right\vert ^{2}-\left[ \left( \nabla \times \mathbf{b}%
\right) \times \mathbf{b}\right] \times \mathbf{b}/\left\vert \mathbf{b}%
\right\vert ^{2}$ and thus 
\begin{eqnarray}
\overline{\left( \nabla \times \mathbf{b}\right) \times \mathbf{u}} &=&
\overline{\frac{\left[ \left( \nabla \times \mathbf{b}\right) \cdot 
\mathbf{b}\right] \mathbf{b}\times \mathbf{u}}{\overline{\left\vert \mathbf{b}
\right\vert ^{2}}}} \nonumber\\
&-&\frac{\overline{\left\{ \left[ \left( \nabla \times 
\mathbf{b}\right) \times \mathbf{b}\right] \times \mathbf{b}\right\} \times 
\mathbf{u}}}{\overline{\left\vert \mathbf{b}\right\vert ^{2}}}  \nonumber \\
&\simeq &-\frac{\overline{\left( \nabla \times \mathbf{b}\right) \cdot 
\mathbf{b}}\;\overline{\mathbf{\mathcal{E}}}}{\overline{\left\vert \mathbf{b}
\right\vert ^{2}}}-\frac{\overline{\left\{ \left[ \left( \nabla \times 
\mathbf{b}\right) \times \mathbf{b}\right] \cdot \mathbf{u}\right\} \mathbf{b
}}}{\overline{\left\vert \mathbf{b}\right\vert ^{2}}}  \nonumber \\
&+&\frac{\overline{\left( \nabla \times \mathbf{b}\right) \times \mathbf{b}}
\;\overline{\mathbf{u}\cdot \mathbf{b}}}{\overline{\left\vert \mathbf{b}
\right\vert ^{2}}}.  \label{apx-r}
\end{eqnarray}%
We obtain for the second term in eq. (\ref{apx-q}): 
\begin{eqnarray}
\overline{\left( \nabla \times \mathbf{b}\right) \times \mathbf{u}}\cdot 
\overline{\mathbf{B}} &\simeq &-\frac{\overline{\left( \nabla \times \mathbf{%
b}\right) \cdot \mathbf{b}}\;\overline{\mathbf{\mathcal{E}}}\cdot \overline{%
\mathbf{B}}}{\overline{\left\vert \mathbf{b}\right\vert ^{2}}}\nonumber\\
&+&\frac{\overline{\mathbf{u}\cdot \mathbf{b}}\;\overline{\left( \nabla \times 
\mathbf{b}\right) \times \mathbf{b}}\cdot \overline{\mathbf{B}}}{\overline{%
\left\vert \mathbf{b}\right\vert ^{2}}}  \nonumber \\
&\simeq &-\frac{k_{S}^{2}H_{S}^{M}\overline{\mathcal{E}}^{\overline{B}}}{%
2E^{b}}+\frac{H_{S}^{C}\overline{F}^{\overline{B}}}{2E^{b}},  \label{apx-s}
\end{eqnarray}%
where the second term in the second row of expr. (\ref{apx-r}) was
considered to give a null contribution when averaged. Defining $%
E_{2}^{b}\equiv \left( E^{b}\right) ^{2}$ we can write the evolution
equation for the small scale magnetic energy as 
\begin{equation}
\frac{\partial E_{2}^{b}}{\partial t}\simeq -2\overline{F}^{\overline{U}}%
\sqrt{E_{2}^{b}}-k_{S}^{2}H_{S}^{M}\overline{\mathcal{E}}^{\overline{B}%
}+H_{S}^{C}\overline{F}^{\overline{B}}-\zeta _{b}E_{2}^{b}.  \label{apx-u}
\end{equation}

\end{document}